\documentclass[reprint,amsmath,amssymb,floatfix,longbibliography]{revtex4-1}

\usepackage{graphicx}           
\usepackage{dcolumn}           % Align table columns on decimal point
\usepackage{bm}                % bold math
\usepackage{subfigure}
\usepackage[normalem]{ulem}    % underlining package
\usepackage{color}             % underlining package
\usepackage{hyperref}          % hyperlink package
\usepackage{natbib}

\newcommand{\BLUE}{black}    % Modification done by myself 

\newcommand{\kT}{k_\textmd{B}T}

\draft % marks overfull lines with a black rule on the right

\begin{document}

\title{Scaling Behaviors of a Polymer Ejected from a Cavity through a Small Pore}

\author{ Hao-Chun Huang$^{1}$ and Pai-Yi Hsiao$^{1,2}$}
\email[Corresponding author, email:\ ]{pyhsiao@mx.nthu.edu.tw}
\address{\small
$^{1}$Department of Engineering and System Science, National Tsing Hua University, Hsinchu, Taiwan, R.~O.~C.\\
$^{2}$Institute of Nuclear Engineering and Science, National Tsing Hua University, Hsinchu, Taiwan, R.~O.~C.}

\date{\today}

\begin{abstract}
Langevin dynamics simulations are performed to investigate ejection dynamics of spherically confined flexible polymers through a pore.
By varying the chain length $N$ and the initial volume fraction $\phi_0$ of the monomers,
two scaling behaviors for the ejection velocity $v$ on the monomer number $m$ in the cavity are obtained: 
$v \sim m^{1.25}\phi_0^{1.25}/N^{1.6}$ for large $m$ and 
$v \sim m^{-1.4}$ as $m$ is small.
A robust scaling theory is developed by dividing the process into the confined and the non-confined stages, and the dynamical equation is derived via the study of energy dissipation.
After trimming the prior stage related to the escape of the head monomer across the pore, the evolution of $m$ is shown to be well described by the scaling theory. 
The ejection time exhibits two proper scaling behaviors: $N^{2/(3\nu)+y_1}\phi_0^{-2/(3\nu)}$ and $N^{2+y_2}$ under the large and small $\phi_0$- or $N$-conditions, respectively, where  $y_1=1/3$, $y_2=1-\nu$, and $\nu$ is the Flory exponent.  
\end{abstract}

\maketitle 
%=============================================================

% Introduction --------------------------------------------------------------------------------------------
Translocation of biopolymers via small pores is a very important biological process;
it allows exchange of large biomolecules, such as DNA, RNA and proteins, between different cellular compartments \cite{alberts2014molecular}. 
When passing through the pores, the conformation of biomolecules is significantly changed in order to fit with the pores which have typically the size of few monomers.   
It creates a large entropic barrier and therefore, driving forces such as chemical potential gradient are generally required to effectuate a translocation \cite{muthukumar2011polymer, milchev2011single, palyulin2014polymer}. 
\textcolor{\BLUE}{In this study, we focus on a special type of driving: polymer translocation induced by spatial confinement.
A vital example is the ejection of a DNA molecule from a virus capsid to a bacteria cell \cite{alberts2014molecular, purohit2005forces}.
Application examples in nanotechnology include trapping single DNA in a nanocage on a membrane \cite{liu2015entropic}, 
transportation of DNA between nanotraps \cite{zhang2018single}, 
gene therapy using engineered protein shells as the transfection vectors \cite{rohovie2017viruslike, edwardson2019virus-inspired}, 
and so on.
These topics require fundamental understanding of packing or ejecting a biopolymer into or from a closed shell.   
}

Muthukumar \cite{muthukumar1999polymer, muthukumar2001translocation} \textcolor{\BLUE}{has studied polymer ejection by considering it as a nucleation problem,} and predicted that the ejection time scales as $\tau\sim N^{1+1/(3\nu)}\phi_0^{-1/(3\nu)}$ where $N$ is the number of monomers, $\phi_0$ is the \textcolor{\BLUE}{volume fraction (vf)} of the monomers prior to ejection, and $\nu$ is the Flory exponent. 
Using the scaling theory and Monte Carlo simulations, Cacciuto and Luijten (CL) \cite{cacciuto2006self, cacciuto2006confinement} argued that the ejection time should be $\tau\sim N^{1+\nu}\phi_0^{-1/(3\nu-1)}$ for a polymer escaped from a spherical cavity.
It was issued from the Kantor and Kardar's expression $\tau\sim N^{1+\nu}/\Delta\mu$ \cite{kantor2004anomalous} by setting the chemical potential difference $\Delta\mu$ to the estimated free energy per monomer $F/N\sim\phi_0^{1/(3\nu-1)}$.
\textcolor{\BLUE}{The exponent $1+\nu$ depicted a lower-bound time scale for the polymer to diffuse unimpededly over its size.}
Sakaue and Yoshinaga (SY) \cite{sakaue2009dynamics} pointed out that $\Delta\mu$ should decrease with the process.
They studied ejection dynamics by balancing the free energy change with the dissipation of the mechanical energy near the pore.
The ejection time was deduced to scale asymptotically as $\tau\sim N^{(2+\nu)/(3\nu)}\phi_0^{-(2+\nu)/(3\nu)}$
at the osmotic driven stage.

\textcolor{\BLUE}{Simulations, on the other hand, revealed that translocation behaviors can be altered by the details of the escape pore \cite{dehaan2010mapping, linna2014dynamics}, the cavity \cite{polson2015polymer, polson2019polymer}, the solvent \cite{ali2008ejection, piili2017uniform}, the chain stiffness \cite{linna2017rigidity, polson2019polymer}, \textit{etc.} 
Despite of the varieties, universalities can be still traced.  
It is thus very important to study scaling physics underpinning the phenomena.  
To have good knowledge on polymer ejection and resolve the non-consistency in the literatures,}
we perform elaborate numerical study in this Letter.
In addition to the ejection time, the variations of the ejection velocity and the monomer number in the cavity during the process are attentively studied.
It permits to rederive the dynamic equation and reveals various astonishing scaling behaviors.

%[2] model and setting-------------------------------------------------------------------
We apply molecular dynamics simulations \cite{plimpton1995fast} to study polymer ejection from a spherical cavity through a small pore to an open semi-space.
The polymer is modeled by a bead-spring chain where the connectivity between beads is described by a harmonic potential with the spring constant $k_{\rm b}=600k_{\rm B}T/\sigma^2$ and the equilibrium bond length $b_0=1.0\sigma$.
The excluded volume of the beads is modeled by the Weeks-Chandler-Andersen potential \cite{weeks1971role} with the parameters $\varepsilon_{\rm m}=1.2k_{\rm B}T$ and $\sigma_{\rm m}=1.0\sigma$.
Here $k_{\rm B}$ is the Boltzmann constant and the temperature $T$ is controlled by Langevin thermostat.
To shorten the notation, the units of all the physical quantities reported latter will not be given in supposing that   
$\mathfrak{m}$ (the mass of a monomer), $\sigma$, $k_{\rm B}T$ are the three basic units of mass, length and energy, respectively.

\begin{figure}[htbp]
	\begin{center}
		\includegraphics[height=0.15\textwidth,angle=0]{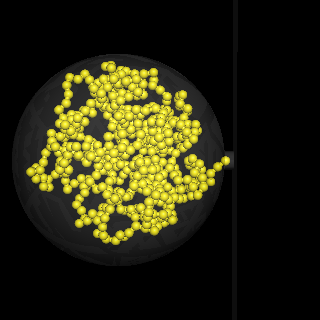}
		\includegraphics[height=0.15\textwidth,angle=0]{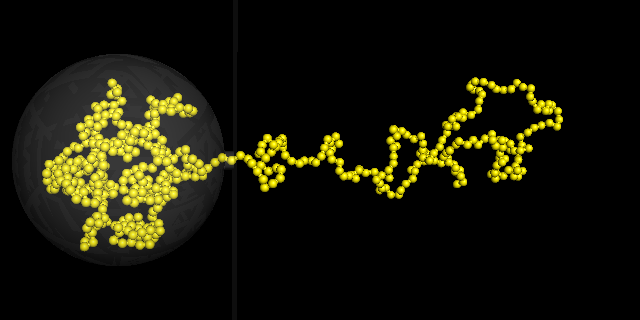}
		\caption{Snapshots of simulation for $N=512$ at $\phi_0=0.2$, at (a) the starting point and (b) $30\%$ of the ejection process.
		}\label{fig:snapshots}
	\end{center}
\end{figure}

We first pump a chain into a cavity and equilibrate the system by constraining the head monomer at the pore entrance as shown in Fig.~\ref{fig:snapshots}(a).
The monomer-cavity wall interaction is modeled by a repulsive Lennard-Jones 9-3 potential with the parameters ($\varepsilon_{\rm w}$, $\sigma_{\rm w}$) = ($3.0$, $1.0$), cut at $r_{\rm c}=\sqrt[6]{2/5}$. 
Various chain lengths, ranged from $N=16$ to $1024$, are investigated.
By varying the cavity diameter $D_{\rm c}$, we are able to study ejection processes at different initial vf, $\phi_0=N(\sigma/D)^3$, where $D=D_{\rm c}-0.5$ is the effective cavity diameter after subtracting the wall thickness. 
The value of $\phi_0$ is varied from $0.4$ down to $0.4\times 2^{-13}$.
The special case $\phi_0=0$ is studied too, which has an infinite $D$ value and corresponds to a translocation through a flat membrane.
The effective pore diameter $d$ is set to $1.5$ and the pore length $\ell$ is $1.25$.

An ejection process is started by removing the constraint of the head monomer.
To obstruct falling of the head monomer into the cavity, particularly when $\phi_0$ is small, 
a wall potential is set at the pore entrance, which acts only on the head monomer and reflects it to the outside.
\textcolor{\BLUE}{We run simulations at different pairs of parameter ($N$, $D$)}.
For each pair, five hundred independent ejection events are studied. 
Snapshots of simulation are illustrated in Fig.~\ref{fig:snapshots}.

\begin{figure}[htbp]
\begin{center}
	\includegraphics[width=0.30\textwidth,angle=270]{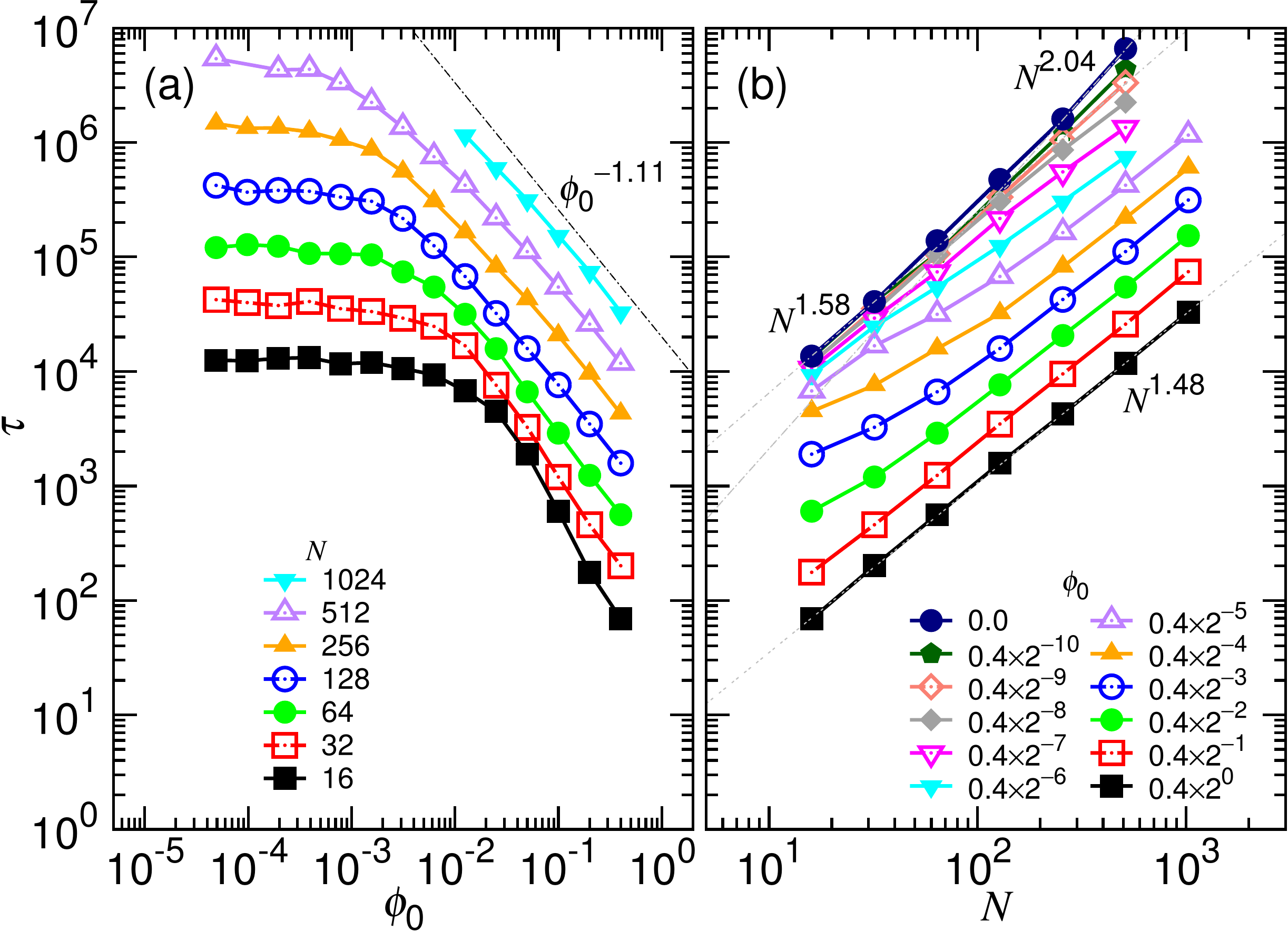}
	\caption{(a) Mean ejection time $\tau$ vs.~the initial volume fraction $\phi_0$ at fixed chain lengths. The $N$ values are given in the legend. (b) $\tau$ vs.~$N$ at fixed $\phi_0$. The values of $\phi_0$ can be read in the figure.
	}\label{fig:tau-tot}
\end{center}
\end{figure}

%[3] tau-tot vs. phi0 and N -------------------------------------------------------------------
We first study the variation of the mean ejection time $\tau$ vs.~the initial vf $\phi_0$ under the fixed chain length condition in Fig.~\ref{fig:tau-tot}(a).
When $\phi_0$ is large, $\tau$ scales approximately as $\phi_0^{-1.11}$. 
It is close to CL's result $\phi_0^{-1/(3\nu-1)}$ \cite{cacciuto2006confinement}. 
However, detailed derivation given later shows that the exponent should be asymptotically $-\frac{2}{3\nu}$.    
Our simulations go further to investigate the small $\phi_0$ behavior and clearly show that the curves are leveled off to a limiting value which is the time required for the polymer to translocate across a flat wall.  

The variation of  $\tau$ vs.~$N$ under the fixed $\phi_0$ condition is studied in Fig.~\ref{fig:tau-tot}(b).
The cases with large $\phi_0$, $0.4$ and $0.2$, show good scaling behaviors $N^{1.48}$, which looks close to the SY's prediction $N^{(2+\nu)/(3\nu)}$ \cite{sakaue2009dynamics}.
At smaller $\phi_0$, the scaling dependence \textcolor{\BLUE}{is not well kept}.
For example, the exponent changes gradually from $1.58$ to $2.04$ as $N$ increases at null $\phi_0$.     

\begin{figure}[htbp]
	\begin{center}
		\includegraphics[width=0.30\textwidth,angle=270]{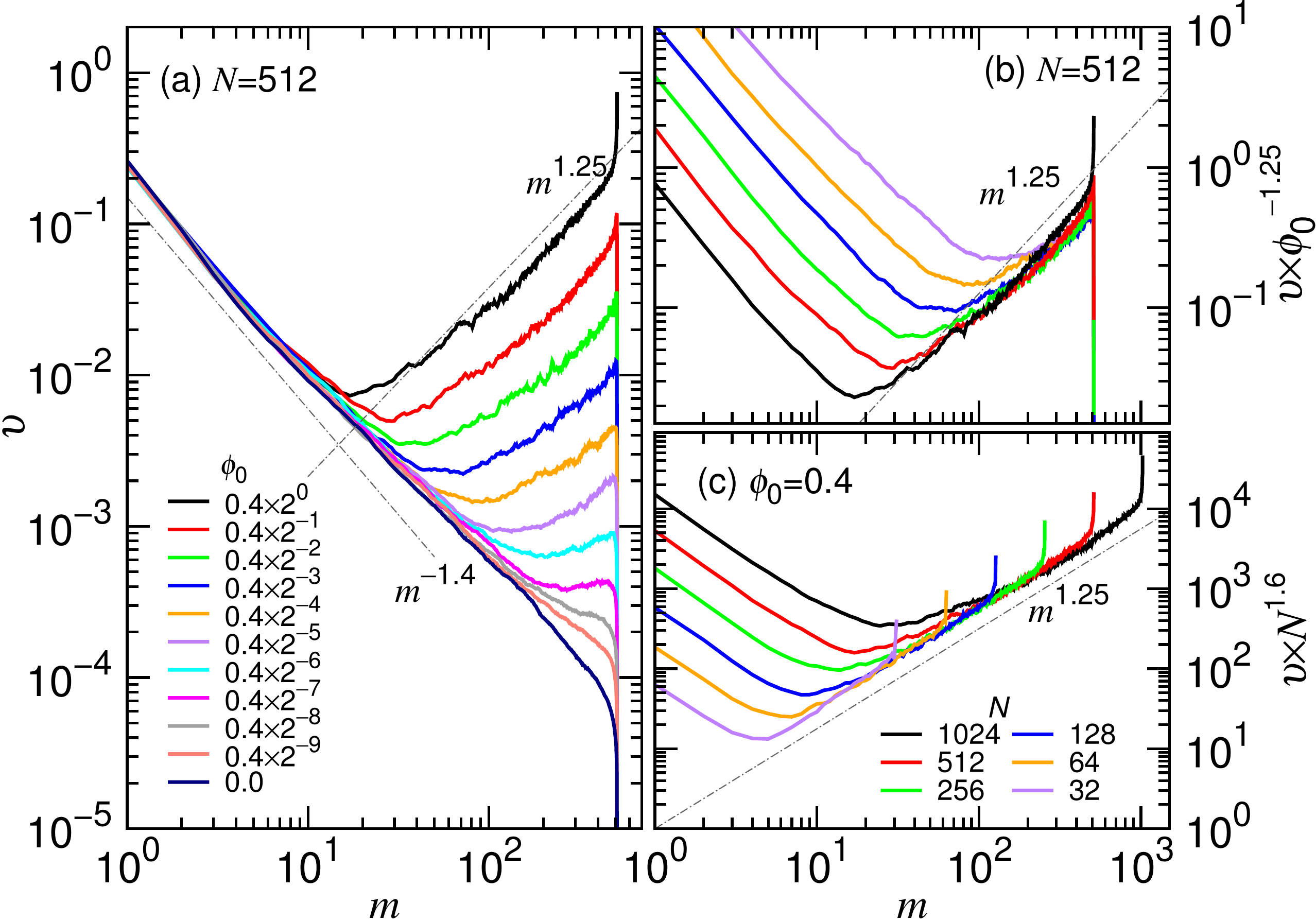}
		\caption{(a) Mean ejection velocity $v$ as a function of the number $m$ of the monomers in the cavity for different $\phi_0$ values where $N=512$. 
		(b) $v\times\phi_0^{-1.25}$ as a function of $m$ at $N=512$.
		The value of $\phi_0$ can be read in the legend (a). 
		(c) $v\times N^{1.6}$ vs.~$m$ with $\phi_0$ being fixed at $0.4$.
		}\label{fig:Vm}
	\end{center}
\end{figure}

To understand the details, the dynamics of ejection is studied. 
We calculate the mean ejection velocity $v$ by taking the reciprocal of the mean dwelling time,
and the results are plotted in Fig.~\ref{fig:Vm}(a) against the number $m$ of the monomers in the cavity.
Two characteristic scalings are discovered.
First, the velocity profile behaves roughly as $m^{1.25}$ when $m$ is large.
With decreasing $m$ below a threshold $m_*$, $v$ turns to show a second scaling $m^{-1.4}$.
Noticeably, $m_*$ depends on $\phi_0$ and the curves are merged together as $m<m_*$  to follow a common tread. 
Further analysis shows that the velocity profiles at the large $m$ section can be collapsed by multiplying $\phi_0^{-1.25}$ under the fixed $N$ condition, as given in Fig.~\ref{fig:Vm}(b).
For varied $N$, Fig.~\ref{fig:Vm}(c) shows that the curves fall on a line at a fixed large $\phi_0$ and $m>m_*$ if $v$ is scaled by $N^{1.6}$.
The results depict two scaling behaviors for the ejection dynamics:
$v \sim m^{1.25}\phi_0^{1.25}/N^{1.6}$ as $m>m_*$ and 
$v \sim m^{-1.4}$ as $m<m_*$.

%[4] scaling theory ------------------------------------------------------------------------------------------------
\color{\BLUE}
Based on the simulations, we develop a scaling theory.
An ejection process can be separated into two stages.
At the first stage (called the confined stage), the instantaneous vf of monomer in the cavity, $\phi=m(\frac{\sigma}{D})^3$, is larger than the overlap vf, $\phi_*$.
The second stage is the non-confined stage, occurred when $\phi$ becomes smaller than $\phi_*$. 
At the separation, the threshold $m_*$ satisfies the condition $D\sim \sigma m_*^{\nu}$. 
It gives $\phi_* \sim (\frac{D}{\sigma})^{-1/(z\nu)}$ where $z=\frac{1}{3\nu-1}$.

At the first stage, the chain portion in the cavity constitutes a semi-dilute solution and builds up excluded volume correlations of length scale $\xi$ (the blob size).
Assume that a blob comprises $g$ monomers.
The space-filling conditions, $\xi\sim\sigma g^{\nu}$ and $g(\frac{\sigma}{\xi})^3\sim \phi$, yield 
$g\sim m^{-z}(\frac{D}{\sigma})^{3z}$.
The free energy is thus given by $F(m) \sim \kT\,m/g$.
Balance the rate of the free energy change, $dF/dt$, with the dissipation rate of energy at the pore, $-\eta v^2$.
The dynamic equation of ejection is deduced,   
$\frac{dm}{dt} \sim -\frac{1}{\Delta\mathfrak{t}}\frac{m^{z}}{(D/\sigma)^{3z}}$.
Here $\eta$ is the friction coefficient, $v=-\frac{\sigma dm}{dt}$ the ejection velocity, and  $\Delta\mathfrak{t}=\frac{\eta\sigma^2}{\kT}$ the time scale of monomer diffusion.
\color{black}
For $\nu=0.6$,  the equation depicts a $m$-scaling behavior with the exponent $z=1.25$, consistent with the results of Fig.~\ref{fig:Vm}.
Substitute $(D/\sigma)^3$ with $N/\phi_0$.
An elimination of the scaling dependence of $v$ on $\phi_0$ can be done by multiplying $\phi_0^{-z}$, as having been shown in Fig.~\ref{fig:Vm}(b).
However, the dependence on $N$ at fixed $\phi_0$ can be only removed by multiplying $N^{1.6}$, not by $N^{z}$,
following the result of  Fig.~\ref{fig:Vm}(c).
It suggests the existence of an extra scaling on $N$ in the equation.
We conjecture that $\eta$ scales additionally with $N$ as $\eta_0 N^{y_1}$ where $y_1 \simeq 0.35$.
The origin of this dependence can be attributed to a combination effect of the \textcolor{\BLUE}{cavity-to-pore space reduction} and the crowding of the monomers from the trans side. 
It slows down the process and as a consequence, a longer chain acquires a smaller velocity when reaching at a state $m$ if $D$ is fixed.
The velocity scaling now reads as 
\begin{equation}
v\sim\frac{\sigma}{\Delta\mathfrak{t}_0}\frac{m^{z}}{N^{y_1}(D/\sigma)^{3z}}.
\end{equation}  

The second stage is commenced by $m < m_*$ \textcolor{\BLUE}{where the chain segments become dilute in the cavity}.
The free energy is about $F(m) \simeq \kT((1-\gamma_{\rm i}')\ln m +(1-\gamma_{\rm o}')\ln (N-m))- m\Delta\mu_{\rm io}$ 
where $\gamma_{\rm i}'$ and $\gamma_{\rm o}'$ describe the scaling of the partition function for a tethered chain inside and outside the cavity, respectively, and $\Delta\mu_{\rm io}$ is the chemical potential difference per monomer \cite{eisenriegler1982adsorption, muthukumar1999polymer, muthukumar2003polymer, muthukumar2011polymer}.
With the help of the rate balance, we obtain $\frac{dm}{dt}\simeq \frac{-1}{\Delta\mathfrak{t}} (\frac{1-\gamma_{\rm i}'}{m}-\frac{1-\gamma_{\rm o}'}{N-m}-\frac{\Delta\mu_{\rm io}}{\kT})$.
\textcolor{\BLUE}{Fig.~\ref{fig:Vm}(a) has revealed an astonishing result that 
$v$ is not sensitive to $\phi_0$ as $m<m_*$ and follows a universal tread of scaling.
It indicates that the $\Delta\mu_{\rm io}$ term is not important.
Moreover, the term related to $\frac{1}{N-m}$ can be omitted too because $N$ is much larger than $m$.
The discrepancy on the resulting exponent for $m$, $-1$, with the simulation, $-1.4$,  
further suggests that the energy dissipation does not come from the sole monomer at the pore.}
The displacing motion of the monomers in the cavity should also participate the dissipation.
The effect can be accounted through a scaling on the effective friction coefficient: $\eta\sim \eta_0 m^{y_2}$.
As a result, the velocity at this stage is described by 
\begin{equation}
v\sim \frac{\sigma}{\Delta\mathfrak{t}_0}m^{-(1+y_2)}
\end{equation}
with the exponent $y_2$ being about $0.4$.

We solve $m(t)$ by integrating the two velocity equations at the three boundary conditions: 
\textcolor{\BLUE}{$m(0)=N$, $m(\tau_1)=m_*$, and $m(\tau_1+\tau_2)=0$.}
Two asymptotic behaviors are obtained.
At the beginning of the ejection, 
\begin{equation}
\frac{m}{N}\simeq \left(1+\frac{t}{t_0}\right)^{-\zeta_1}
\label{eq:confined}
\end{equation}
and near the end of the ejection,
\begin{equation}
\frac{m}{N}\simeq \left(1-\frac{t}{\tau_1+\tau_2}\right)^{\zeta_2}
\label{eq:non-confined}
\end{equation}
where $\zeta_1=\frac{1}{z-1}$, $\zeta_2=\frac{1}{2+y_2}$, and $t_0=\frac{N^{1+y_1}\phi_0^{-z} \Delta\mathfrak{t}_0}{z-1}$.
We remark that the confined stage is not always involved in an ejection process.
When $\phi_0<\phi_0^*\sim N^{-1/z}$, the pervaded space of the entire chain is smaller than the cavity.
The system is processed only through the non-confined stage.
  
The total ejection time are calculated.
For the case with fixed chain length,  $\tau_1+\tau_2$ varies with $\phi_0$ as 
\begin{eqnarray}
\begin{cases}
\frac{\Delta\mathfrak{t}_0}{2+y_2}N^{2+y_2}, & \phi_0 \le \phi_0^* \\[7pt]
\frac{\Delta\mathfrak{t}_0 N^{y_1}}{z-1}\left[\left(\frac{N}{\phi_0}\right)^{\frac{2}{3\nu}}-\frac{N}{\phi_0^{z}}\right] + \frac{\Delta\mathfrak{t}_0}{2+y_2}\left(\frac{N}{\phi_0}\right)^{\frac{2+y_2}{3\nu}}, & \phi_0>\phi_0^* \\
\end{cases}\label{eq:tau_vs_phi0_N}
\end{eqnarray}
If $N$ is varied and $\phi_0$ is fixed, the resulting expression is the same but the conditions $\phi_0 \le \phi_0^*$ and $\phi_0 > \phi_0^*$ are replaced  by $N\le N_*$ and $N>N_*$, respectively,  where $N_*\sim \phi_0^{-z}$.
The results show that the ejection time is a sum of several terms if the confined stage is involved.
For large $N$ and $\phi_0$, the dominated term is $N^{\frac{2}{3\nu}+y_1}\phi_0^{-\frac{2}{3\nu}}$.
By taking $\nu=0.6$ and $y_1=0.35$, the ejection time has the scaling $N^{1.46} \phi_0^{-1.11}$.
This is exactly what we have observed in Figs.~\ref{fig:tau-tot}(a) and \ref{fig:tau-tot}(b), when both $\phi_0$ and $N$ are large.
Eq.~\ref{eq:tau_vs_phi0_N} also predicts the leveling-off of the ejection time to a value around $N^{2+y_2}$ as $\phi_0<\phi_0^*$ under the fixed-$N$ condition.
However, discrepancy is found with the simulations. 
The predicted $N^{2.4}$ behavior (by setting $y_2=0.4$) is not clearly seen in Fig.~\ref{fig:tau-tot}(b). 
For example, the exponent changed gradually from $1.58$ to $2.04$ at $\phi_0=0$.
Could it be a result of a strong finite-size effect of the chain length?

\begin{figure}[htbp]
	\begin{center}
		\includegraphics[width=0.30\textwidth,angle=270]{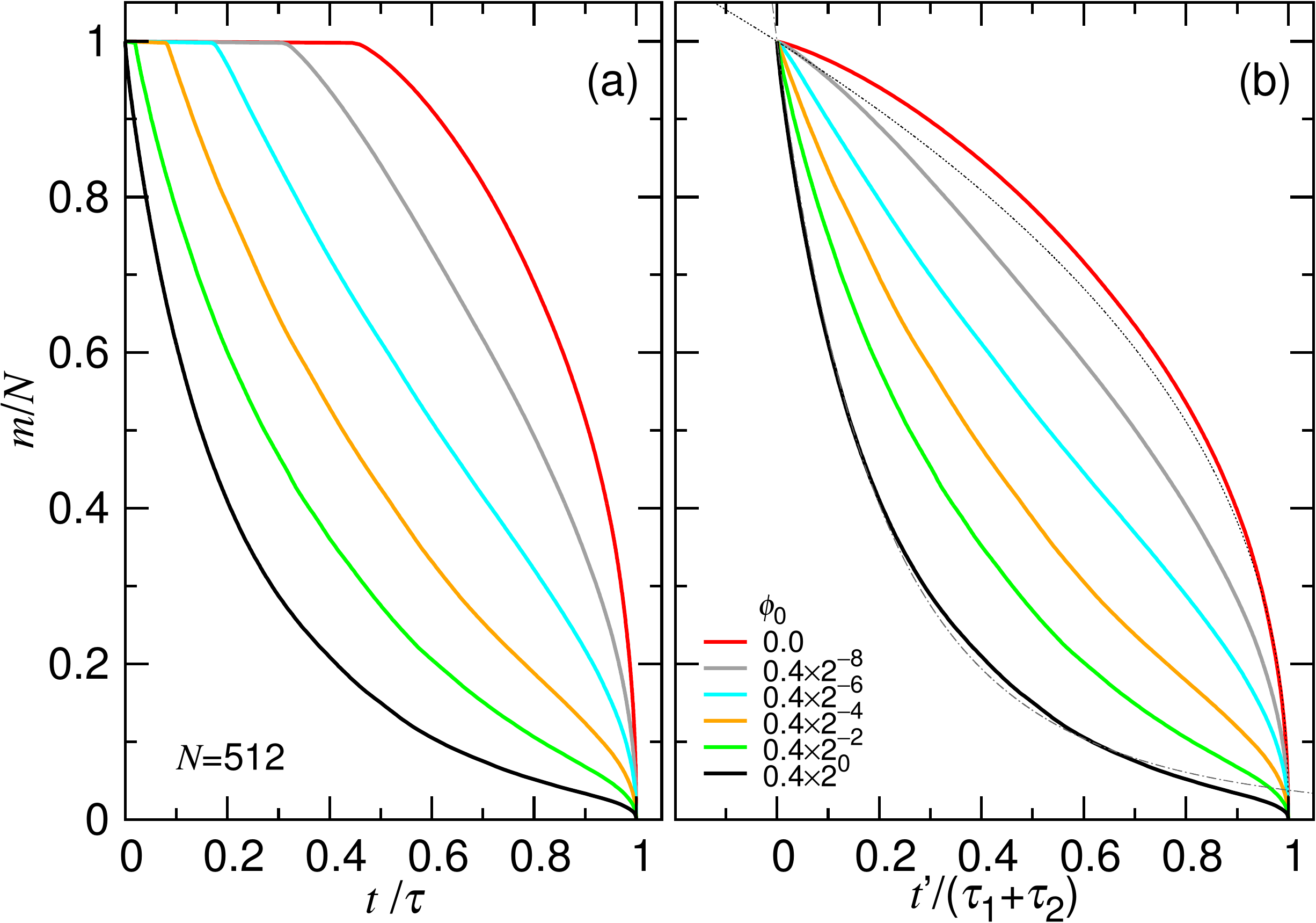}
		\caption{(a) $m/N$ vs.~$t/\tau$ and (b) $m/N$ vs.~$t'/(\tau_1+\tau_2)$, for $N=512$ at different $\phi_0$.
			     The value of $\phi_0$ can be read in the legend.
			     $t'$ and $\tau_1+\tau_2$ are, respectively, the elapsed time and the ejection time after trimming off the prior stage. 
		}\label{fig:m_vs_t}
	\end{center}
\end{figure}

To answer the question, we investigate the evolution of the mean monomer number in the cavity.
The normalized variations, $m/N$ vs.~$t/\tau$, for various $\phi_0$ at $N=512$ are plotted in Fig.\ref{fig:m_vs_t}(a).
The curves exhibit a plateau structure at the beginning of the process and change to show smooth decreasing some moment later.  
The smaller the initial vf, the wider the plateau.
We found that the changing point occurred at the moment when the head monomer left the pore, \textit{i.e.}~at $m=N-2$ in this study.
And the time spent in the plateau region can be as long as $45\%$ of the total ejection time, as shown in the figure, and this is evidently not negligible.
It turns out clear that there exists a stage, prior to the confined or the non-confined stages, for the head monomer to find a way out of the pore.  
The time spent in this stage is denoted by $\tau_0$.
Therefore, the total ejection time is $\tau=\tau_0+\tau_1+\tau_2$. 

\textcolor{\BLUE}{We trim the prior stage, and plot $m/N$ vs.~$t'/(\tau_1+\tau_2)$ in Fig.\ref{fig:m_vs_t}(b)} where $t'=t-\tau_0$.
We are now at the position to examine the asymptotic behavior given by Eq.~\ref{eq:confined}.
The variation curve for $\phi_0=0.4$ was fit by using \textcolor{\BLUE}{Levenberg--Marquardt method \cite{bates1988nonlinear}, with $\zeta_1$ set to $4.0$ and $t_0$ being the fitting parameter}.
The fitting curve has been plotted in dash-dotted line.
Good agreement with the data is found. 
To verify Eq.~\ref{eq:non-confined}, we chose the extreme case $\phi_0=0.0$ by setting $\zeta_2=2.4^{-1}\simeq 0.417$ and plotted the equation in dotted line.
The consistency is satisfactory as $m$ is small.
Therefore, the dynamics of ejection after the prior stage can be well described by the pictures of the two-stage model.

\begin{figure}[htbp]
	\begin{center}
		\includegraphics[width=0.30\textwidth,angle=270]{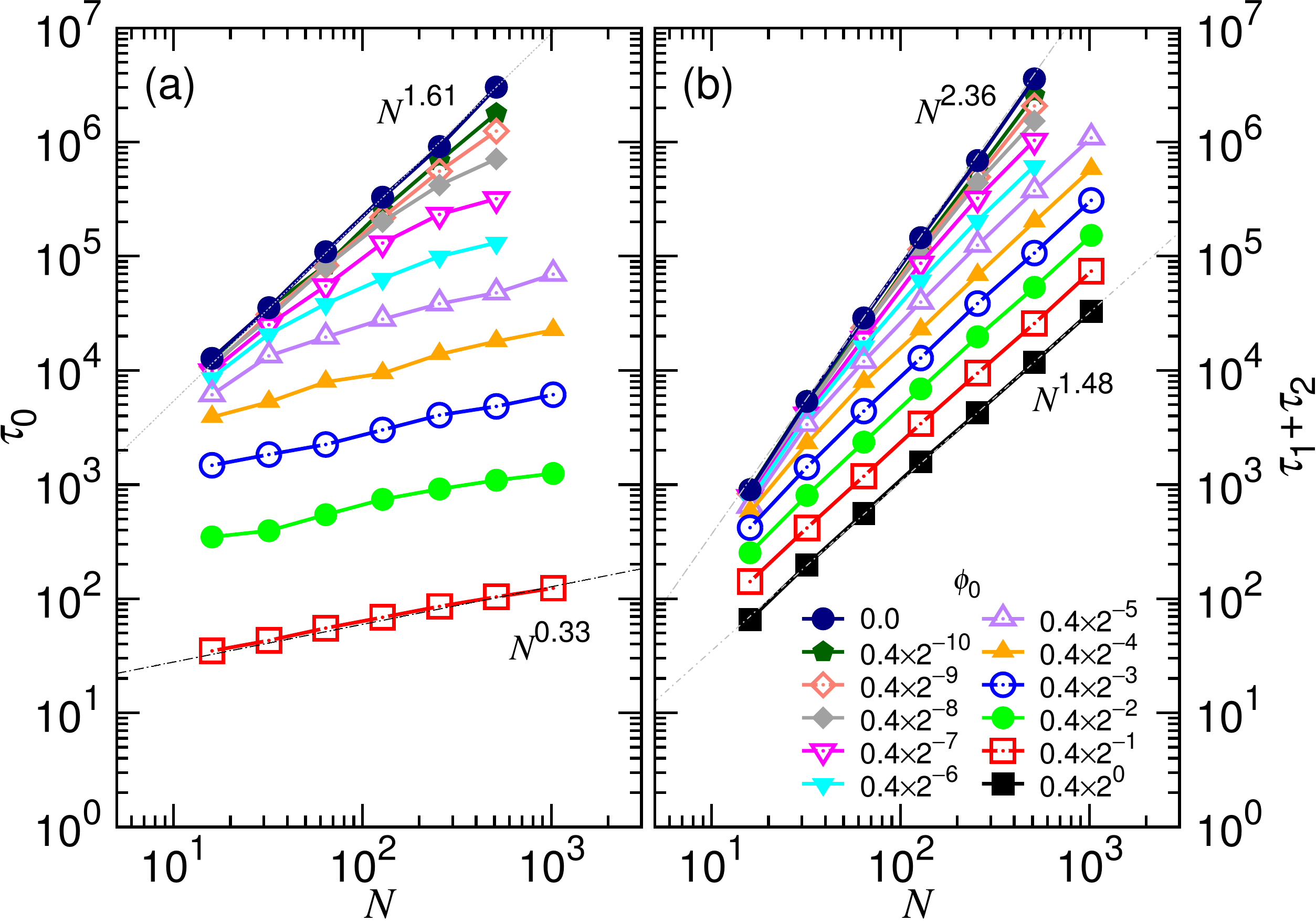}
		\caption{(a) $\tau_0$ vs.~$N$ and (b) $\tau_1+\tau_2$ vs.~$N$, at different $\phi_0$. 
			The values of $\phi_0$ can be read in the legend of (b).
		}\label{fig:tau-decomp}
	\end{center}
\end{figure}

We further decompose the ejection time into two components, $\tau_0$ and $\tau_1+\tau_2$, and study their variations against $N$ in Fig.~\ref{fig:tau-decomp}. 
We find that $\tau_0$ follows two scaling behaviors.
At large $\phi_0$  such as $0.2$, the exponent is about $0.33$.
Decreasing $\phi_0$ slows down the process and moves upward the $\tau_0$ curves in a parallel way.
Notably, $\tau_0$ has an upper bound occurred at null $\phi_0$.
At the bound,  it follows a second scaling with the exponent equal to $1.61$.
Thus, the curves are deflected gradually to the larger scaling if $N<N_*$ is met, as seen in the figure.

We remark that the prior stage does not appear at $\phi_0=0.4$.
\textcolor{\BLUE}{It is because the value is larger than the vf inside the pore, estimated by $\phi_{\rm p} = \frac{1}{6}\pi\sigma^3/(\frac{1}{4}\pi d^2 \sigma) \simeq 0.296$.	
The ejection thus started in an imminent way, driven by the osmotic pressure difference from the cavity.}
For the cases with $\phi_0<\phi_p$, the head monomer needs to overcome an energy barrier when crossing the pore.
We have verified that the prior time grows exponentially with the pore length $\ell$.
It showed that the prior stage is, in fact, a Kramers escape problem and  $\tau_0$ is described by $ \eta\exp( \frac{\Delta\mu_{\rm cp} \ell}{\kT\sigma})$ where $\Delta\mu_{\rm cp}$ is the chemical potential difference between the cis and the pore region \cite{kramers1940brownian, gardiner2004handbook}.
At large $\phi_0$ such as $0.2$, the \textcolor{\BLUE}{cavity-to-pore space shrinkage} imposes a $N$-scaling on the friction coefficient with  $y_1=1/3$.
Consequently, $\tau_0$ scales as $N^{1/3}$, consistent with the observations. 
For $\phi_0<\phi_0^*$, the effective friction of the chain can be shown proportional to $N$ times a factor $N^{\nu}$ which accounts the slowing down due to the \textcolor{\BLUE}{spatial reduction} in transporting the chain coil from the cis region into the pore.
It explains the small-$\phi_0$ behavior with a scaling exponent equal to $1+\nu$.
\textcolor{\BLUE}{In this situation, the solution is dilute and the $m$ monomers on the cis side participate
the energy dissipation with a displacing velocity  of about $v/m^{\nu}$	
The dissipation rate $\eta v^2$ is thus replaced by $ m\cdot\eta_0 m^{\nu}\cdot(v/m^{\nu})^2$.
It gives $y_2=1-\nu$.}
   
Fig.~\ref{fig:tau-decomp}(b) shows $\tau_1+\tau_2$ vs.~$N$ at various $\phi_0$.
Compared with the total ejection time in Fig.~\ref{fig:tau-tot}(b), the scaling behavior becomes neater.
The exponent is $1.48$ at $\phi_0=0.4$ and the curves upshift with decreasing $\phi_0$ in a parallel way in the log-log plot.
Similar to $\tau_0$, the upshifted curves are bent to follow a second scaling $N^{2.36}$ when $N<N_*$. 
The results are consistent  with the prediction of Eq.~\ref{eq:tau_vs_phi0_N} where 
the dominated scaling switches from $\frac{2}{3\nu}+y_1$ to $2+y_2$ as $N$ decreases.
We have also examined the decomposition of $\tau$ vs.~$\phi_0$.
Neater scaling results consistent with the prediction were observed.
\textcolor{\BLUE}{Since $\tau_0$ is small at large $\phi_0$, 
the $\tau_1+\tau_2$ curve retains the scaling exponent $-\frac{2}{3\nu}$.
The exponent changes to $0$ as $\phi_0 < \phi_0^*$, similar to the behavior in Fig.~\ref{fig:tau-tot}(a).}

% [Conclusion]---------------------------------------------------------------------------------------------
It is now clear that an ejection process comprises mainly three stages: the prior stage, the confined stage and the non-confined stage.
The above study has shown that the contribution from the prior stage is significant and can affect the scaling calculations for the total ejection time.
This might explain why non-consistent scaling results were obtained in literatures \cite{muthukumar2001translocation, cacciuto2006self, cacciuto2006confinement, sakaue2009dynamics, milchev2011single, palyulin2014polymer}.
After trimming the prior stage, we obtained the proper ejection scalings: $\tau\sim N^{2/(3\nu)+y_1}\phi_0^{-2/(3\nu)}$ for $\phi_0>\phi_0^*$ or $N>N_*$ and  $N^{2+y_2}$ for $\phi_0<\phi_0^*$ or $N<N_*$ where
$y_1=1/3$ and $y_2=1-\nu$.
A robust scaling theory has been developed.
The ejection dynamics, including the evolution of the velocity and the number of monomers in the cavity, support the physical pictures depicted by the theory.

\color{\BLUE}
A number of open questions can be studied in the future.
Notably, a comprehensive understanding of the velocity profile will be  helpful.
For example, the velocity curve is somewhat bent up at the confined stage.
It can be attributed to the virial coefficient effect in the osmotic pressure due from the highly-concentrated polymer solutions \cite{paturej2019universal}. 
Packing or ejecting a chain depends very much on the bending rigidity.
How the velocity profile and scaling classes are altered by the chain rigidity is urgent to be known.
The current work treated the problem without considering hydrodynamic interaction.
This topics is expected to play a significant role and the influences on the ejection velocity should be understood.
Nevertheless, a simple and clear theory has been provided in this Letter to explain the essential processes of the escape dynamics for confined polymers.
\color{black}

%\section*{Acknowledgments}%============================================================
This material is based upon work supported by the Ministry of Science and Technology, Taiwan under the Contract No.~MOST 106-2112-M-007-027-MY3.

%\bibliographystyle{hsiao-plainnat}
%\bibliography{PETrans_refs}

\begin{thebibliography}{30}%
	\makeatletter
	\providecommand \@ifxundefined [1]{%
		\@ifx{#1\undefined}
	}%
	\providecommand \@ifnum [1]{%
		\ifnum #1\expandafter \@firstoftwo
		\else \expandafter \@secondoftwo
		\fi
	}%
	\providecommand \@ifx [1]{%
		\ifx #1\expandafter \@firstoftwo
		\else \expandafter \@secondoftwo
		\fi
	}%
	\providecommand \natexlab [1]{#1}%
	\providecommand \enquote  [1]{``#1''}%
	\providecommand \bibnamefont  [1]{#1}%
	\providecommand \bibfnamefont [1]{#1}%
	\providecommand \citenamefont [1]{#1}%
	\providecommand \href@noop [0]{\@secondoftwo}%
	\providecommand \href [0]{\begingroup \@sanitize@url \@href}%
	\providecommand \@href[1]{\@@startlink{#1}\@@href}%
	\providecommand \@@href[1]{\endgroup#1\@@endlink}%
	\providecommand \@sanitize@url [0]{\catcode `\\12\catcode `\$12\catcode
		`\&12\catcode `\#12\catcode `\^12\catcode `\_12\catcode `\%12\relax}%
	\providecommand \@@startlink[1]{}%
	\providecommand \@@endlink[0]{}%
	\providecommand \url  [0]{\begingroup\@sanitize@url \@url }%
	\providecommand \@url [1]{\endgroup\@href {#1}{\urlprefix }}%
	\providecommand \urlprefix  [0]{URL }%
	\providecommand \Eprint [0]{\href }%
	\providecommand \doibase [0]{http://dx.doi.org/}%
	\providecommand \selectlanguage [0]{\@gobble}%
	\providecommand \bibinfo  [0]{\@secondoftwo}%
	\providecommand \bibfield  [0]{\@secondoftwo}%
	\providecommand \translation [1]{[#1]}%
	\providecommand \BibitemOpen [0]{}%
	\providecommand \bibitemStop [0]{}%
	\providecommand \bibitemNoStop [0]{.\EOS\space}%
	\providecommand \EOS [0]{\spacefactor3000\relax}%
	\providecommand \BibitemShut  [1]{\csname bibitem#1\endcsname}%
	\let\auto@bib@innerbib\@empty
	%</preamble>
	\bibitem [{\citenamefont {Alberts}\ \emph {et~al.}(2014)\citenamefont
		{Alberts}, \citenamefont {Johnson}, \citenamefont {Lewis}, \citenamefont
		{Morgan}, \citenamefont {Raff}, \citenamefont {Roberts},\ and\ \citenamefont
		{Walter}}]{alberts2014molecular}%
	\BibitemOpen
	\bibfield  {author} {\bibinfo {author} {\bibfnamefont {B.}~\bibnamefont
			{Alberts}}, \bibinfo {author} {\bibfnamefont {A.}~\bibnamefont {Johnson}},
		\bibinfo {author} {\bibfnamefont {J.}~\bibnamefont {Lewis}}, \bibinfo
		{author} {\bibfnamefont {D.}~\bibnamefont {Morgan}}, \bibinfo {author}
		{\bibfnamefont {M.}~\bibnamefont {Raff}}, \bibinfo {author} {\bibfnamefont
			{K.}~\bibnamefont {Roberts}}, \ and\ \bibinfo {author} {\bibfnamefont
			{P.}~\bibnamefont {Walter}},\ }\href@noop {} {\emph {\bibinfo {title}
			{Molecular Biology of the Cell}}},\ \bibinfo {edition} {6th}\ ed.\ (\bibinfo
	{publisher} {Taylor \& Francis Group},\ \bibinfo {year} {2014})\BibitemShut
	{NoStop}%
	\bibitem [{\citenamefont {Muthukumar}(2011)}]{muthukumar2011polymer}%
	\BibitemOpen
	\bibfield  {author} {\bibinfo {author} {\bibfnamefont {Murugappan}\
			\bibnamefont {Muthukumar}},\ }\href@noop {} {\emph {\bibinfo {title} {Polymer
				translocation}}}\ (\bibinfo  {publisher} {CRC Press},\ \bibinfo {year}
	{2011})\BibitemShut {NoStop}%
	\bibitem [{\citenamefont {Milchev}(2011)}]{milchev2011single}%
	\BibitemOpen
	\bibfield  {author} {\bibinfo {author} {\bibfnamefont {Andrey}\ \bibnamefont
			{Milchev}},\ }\bibfield  {title} {\enquote {\bibinfo {title} {Single-polymer
				dynamics under constraints: scaling theory and computer experiment},}\
	}\href@noop {} {\bibfield  {journal} {\bibinfo  {journal} {J. Phys. Condens.
				Matter}\ }\textbf {\bibinfo {volume} {23}},\ \bibinfo {pages} {103101}
		(\bibinfo {year} {2011})}\BibitemShut {NoStop}%
	\bibitem [{\citenamefont {Palyulin}\ \emph {et~al.}(2014)\citenamefont
		{Palyulin}, \citenamefont {Ala-Nissila},\ and\ \citenamefont
		{Metzler}}]{palyulin2014polymer}%
	\BibitemOpen
	\bibfield  {author} {\bibinfo {author} {\bibfnamefont {Vladimir~V}\
			\bibnamefont {Palyulin}}, \bibinfo {author} {\bibfnamefont {Tapio}\
			\bibnamefont {Ala-Nissila}}, \ and\ \bibinfo {author} {\bibfnamefont {Ralf}\
			\bibnamefont {Metzler}},\ }\bibfield  {title} {\enquote {\bibinfo {title}
			{Polymer translocation: the first two decades and the recent
				diversification},}\ }\href@noop {} {\bibfield  {journal} {\bibinfo  {journal}
			{Soft matter}\ }\textbf {\bibinfo {volume} {10}},\ \bibinfo {pages} {9016}
		(\bibinfo {year} {2014})}\BibitemShut {NoStop}%
	\bibitem [{\citenamefont {Purohit}\ \emph {et~al.}(2005)\citenamefont
		{Purohit}, \citenamefont {Inamdar}, \citenamefont {Grayson}, \citenamefont
		{Squires}, \citenamefont {Kondev},\ and\ \citenamefont
		{Phillips}}]{purohit2005forces}%
	\BibitemOpen
	\bibfield  {author} {\bibinfo {author} {\bibfnamefont {Prashant~K.}\
			\bibnamefont {Purohit}}, \bibinfo {author} {\bibfnamefont {Mandar~M.}\
			\bibnamefont {Inamdar}}, \bibinfo {author} {\bibfnamefont {Paul~D.}\
			\bibnamefont {Grayson}}, \bibinfo {author} {\bibfnamefont {Todd~M.}\
			\bibnamefont {Squires}}, \bibinfo {author} {\bibfnamefont {Jan{\'{e}}}\
			\bibnamefont {Kondev}}, \ and\ \bibinfo {author} {\bibfnamefont {Rob}\
			\bibnamefont {Phillips}},\ }\bibfield  {title} {\enquote {\bibinfo {title}
			{Forces during bacteriophage {DNA} packaging and ejection},}\ }\href
	{\doibase 10.1529/biophysj.104.047134} {\bibfield  {journal} {\bibinfo
			{journal} {Biophys. J.}\ }\textbf {\bibinfo {volume} {88}},\ \bibinfo {pages}
		{851--866} (\bibinfo {year} {2005})}\BibitemShut {NoStop}%
	\bibitem [{\citenamefont {Liu}\ \emph {et~al.}(2015)\citenamefont {Liu},
		\citenamefont {Skanata},\ and\ \citenamefont {Stein}}]{liu2015entropic}%
	\BibitemOpen
	\bibfield  {author} {\bibinfo {author} {\bibfnamefont {Xu}~\bibnamefont
			{Liu}}, \bibinfo {author} {\bibfnamefont {Mirna~Mihovilovic}\ \bibnamefont
			{Skanata}}, \ and\ \bibinfo {author} {\bibfnamefont {Derek}\ \bibnamefont
			{Stein}},\ }\bibfield  {title} {\enquote {\bibinfo {title} {Entropic cages
				for trapping {DNA} near a nanopore},}\ }\href {\doibase 10.1038/ncomms7222}
	{\bibfield  {journal} {\bibinfo  {journal} {Nature Communications}\ }\textbf
		{\bibinfo {volume} {6}} (\bibinfo {year} {2015}),\
		10.1038/ncomms7222}\BibitemShut {NoStop}%
	\bibitem [{\citenamefont {Zhang}\ \emph {et~al.}(2018)\citenamefont {Zhang},
		\citenamefont {Liu}, \citenamefont {Zhao}, \citenamefont {Yu}, \citenamefont
		{Reisner},\ and\ \citenamefont {Dunbar}}]{zhang2018single}%
	\BibitemOpen
	\bibfield  {author} {\bibinfo {author} {\bibfnamefont {Yuning}\ \bibnamefont
			{Zhang}}, \bibinfo {author} {\bibfnamefont {Xu}~\bibnamefont {Liu}}, \bibinfo
		{author} {\bibfnamefont {Yanan}\ \bibnamefont {Zhao}}, \bibinfo {author}
		{\bibfnamefont {Jen-Kan}\ \bibnamefont {Yu}}, \bibinfo {author}
		{\bibfnamefont {Walter}\ \bibnamefont {Reisner}}, \ and\ \bibinfo {author}
		{\bibfnamefont {William~B.}\ \bibnamefont {Dunbar}},\ }\bibfield  {title}
	{\enquote {\bibinfo {title} {Single molecule {DNA} resensing using a two-pore
				device},}\ }\href {\doibase 10.1002/smll.201801890} {\bibfield  {journal}
		{\bibinfo  {journal} {Small}\ }\textbf {\bibinfo {volume} {14}},\ \bibinfo
		{pages} {1801890} (\bibinfo {year} {2018})}\BibitemShut {NoStop}%
	\bibitem [{\citenamefont {Rohovie}\ \emph {et~al.}(2017)\citenamefont
		{Rohovie}, \citenamefont {Nagasawa},\ and\ \citenamefont
		{Swartz}}]{rohovie2017viruslike}%
	\BibitemOpen
	\bibfield  {author} {\bibinfo {author} {\bibfnamefont {Marcus~J.}\
			\bibnamefont {Rohovie}}, \bibinfo {author} {\bibfnamefont {Maya}\
			\bibnamefont {Nagasawa}}, \ and\ \bibinfo {author} {\bibfnamefont {James~R.}\
			\bibnamefont {Swartz}},\ }\bibfield  {title} {\enquote {\bibinfo {title}
			{Virus-like particles: Next-generation nanoparticles for targeted therapeutic
				delivery},}\ }\href {\doibase 10.1002/btm2.10049} {\bibfield  {journal}
		{\bibinfo  {journal} {Bioeng. Transl. Med.}\ }\textbf {\bibinfo {volume}
			{2}},\ \bibinfo {pages} {43--57} (\bibinfo {year} {2017})}\BibitemShut
	{NoStop}%
	\bibitem [{\citenamefont {Edwardson}\ and\ \citenamefont
		{Hilvert}(2019)}]{edwardson2019virus-inspired}%
	\BibitemOpen
	\bibfield  {author} {\bibinfo {author} {\bibfnamefont {Thomas G.~W.}\
			\bibnamefont {Edwardson}}\ and\ \bibinfo {author} {\bibfnamefont {Donald}\
			\bibnamefont {Hilvert}},\ }\bibfield  {title} {\enquote {\bibinfo {title}
			{Virus-inspired function in engineered protein cages},}\ }\href {\doibase
		10.1021/jacs.9b03705} {\bibfield  {journal} {\bibinfo  {journal} {Journal of
				the American Chemical Society}\ }\textbf {\bibinfo {volume} {141}},\ \bibinfo
		{pages} {9432--9443} (\bibinfo {year} {2019})}\BibitemShut {NoStop}%
	\bibitem [{\citenamefont {Muthukumar}(1999)}]{muthukumar1999polymer}%
	\BibitemOpen
	\bibfield  {author} {\bibinfo {author} {\bibfnamefont {Murugappan}\
			\bibnamefont {Muthukumar}},\ }\bibfield  {title} {\enquote {\bibinfo {title}
			{Polymer translocation through a hole},}\ }\href@noop {} {\bibfield
		{journal} {\bibinfo  {journal} {J. Chem. Phys.}\ }\textbf {\bibinfo {volume}
			{111}},\ \bibinfo {pages} {10371} (\bibinfo {year} {1999})}\BibitemShut
	{NoStop}%
	\bibitem [{\citenamefont {Muthukumar}(2001)}]{muthukumar2001translocation}%
	\BibitemOpen
	\bibfield  {author} {\bibinfo {author} {\bibfnamefont {M.}~\bibnamefont
			{Muthukumar}},\ }\bibfield  {title} {\enquote {\bibinfo {title}
			{Translocation of a confined polymer through a hole},}\ }\href {\doibase
		10.1103/physrevlett.86.3188} {\bibfield  {journal} {\bibinfo  {journal}
			{Phys. Rev. Lett.}\ }\textbf {\bibinfo {volume} {86}},\ \bibinfo {pages}
		{3188--3191} (\bibinfo {year} {2001})}\BibitemShut {NoStop}%
	\bibitem [{\citenamefont {Cacciuto}\ and\ \citenamefont
		{Luijten}(2006{\natexlab{a}})}]{cacciuto2006self}%
	\BibitemOpen
	\bibfield  {author} {\bibinfo {author} {\bibfnamefont {Angelo}\ \bibnamefont
			{Cacciuto}}\ and\ \bibinfo {author} {\bibfnamefont {Erik}\ \bibnamefont
			{Luijten}},\ }\bibfield  {title} {\enquote {\bibinfo {title} {Self-avoiding
				flexible polymers under spherical confinement},}\ }\href {\doibase
		10.1021/nl052351n} {\bibfield  {journal} {\bibinfo  {journal} {Nano Lett.}\
		}\textbf {\bibinfo {volume} {6}},\ \bibinfo {pages} {901--905} (\bibinfo
		{year} {2006}{\natexlab{a}})}\BibitemShut {NoStop}%
	\bibitem [{\citenamefont {Cacciuto}\ and\ \citenamefont
		{Luijten}(2006{\natexlab{b}})}]{cacciuto2006confinement}%
	\BibitemOpen
	\bibfield  {author} {\bibinfo {author} {\bibfnamefont {Angelo}\ \bibnamefont
			{Cacciuto}}\ and\ \bibinfo {author} {\bibfnamefont {Erik}\ \bibnamefont
			{Luijten}},\ }\bibfield  {title} {\enquote {\bibinfo {title}
			{Confinement-driven translocation of a flexible polymer},}\ }\href@noop {}
	{\bibfield  {journal} {\bibinfo  {journal} {Phys. Rev. Lett.}\ }\textbf
		{\bibinfo {volume} {96}},\ \bibinfo {pages} {238104} (\bibinfo {year}
		{2006}{\natexlab{b}})}\BibitemShut {NoStop}%
	\bibitem [{\citenamefont {Kantor}\ and\ \citenamefont
		{Kardar}(2004)}]{kantor2004anomalous}%
	\BibitemOpen
	\bibfield  {author} {\bibinfo {author} {\bibfnamefont {Yacov}\ \bibnamefont
			{Kantor}}\ and\ \bibinfo {author} {\bibfnamefont {Mehran}\ \bibnamefont
			{Kardar}},\ }\bibfield  {title} {\enquote {\bibinfo {title} {Anomalous
				dynamics of forced translocation},}\ }\href@noop {} {\bibfield  {journal}
		{\bibinfo  {journal} {Phys. Rev. E}\ }\textbf {\bibinfo {volume} {69}},\
		\bibinfo {pages} {021806} (\bibinfo {year} {2004})}\BibitemShut {NoStop}%
	\bibitem [{\citenamefont {Sakaue}\ and\ \citenamefont
		{Yoshinaga}(2009)}]{sakaue2009dynamics}%
	\BibitemOpen
	\bibfield  {author} {\bibinfo {author} {\bibfnamefont {Takahiro}\
			\bibnamefont {Sakaue}}\ and\ \bibinfo {author} {\bibfnamefont {Natsuhiko}\
			\bibnamefont {Yoshinaga}},\ }\bibfield  {title} {\enquote {\bibinfo {title}
			{Dynamics of polymer decompression: Expansion, unfolding, and ejection},}\
	}\href@noop {} {\bibfield  {journal} {\bibinfo  {journal} {Phys. Rev. Lett.}\
		}\textbf {\bibinfo {volume} {102}} (\bibinfo {year} {2009})}\BibitemShut
	{NoStop}%
	\bibitem [{\citenamefont {de~Haan}\ and\ \citenamefont
		{Slater}(2010)}]{dehaan2010mapping}%
	\BibitemOpen
	\bibfield  {author} {\bibinfo {author} {\bibfnamefont {Hendrick~W.}\
			\bibnamefont {de~Haan}}\ and\ \bibinfo {author} {\bibfnamefont {Gary~W.}\
			\bibnamefont {Slater}},\ }\bibfield  {title} {\enquote {\bibinfo {title}
			{Mapping the variation of the translocation alpha scaling exponent with
				nanopore width},}\ }\href {https://doi.org/10.1103/physreve.81.051802}
	{\bibfield  {journal} {\bibinfo  {journal} {Phys. Rev. E}\ }\textbf {\bibinfo
			{volume} {81}} (\bibinfo {year} {2010})}\BibitemShut {NoStop}%
	\bibitem [{\citenamefont {Linna}\ \emph {et~al.}(2014)\citenamefont {Linna},
		\citenamefont {Moisio}, \citenamefont {Suhonen},\ and\ \citenamefont
		{Kaski}}]{linna2014dynamics}%
	\BibitemOpen
	\bibfield  {author} {\bibinfo {author} {\bibfnamefont {R.~P.}\ \bibnamefont
			{Linna}}, \bibinfo {author} {\bibfnamefont {J.~E.}\ \bibnamefont {Moisio}},
		\bibinfo {author} {\bibfnamefont {P.~M.}\ \bibnamefont {Suhonen}}, \ and\
		\bibinfo {author} {\bibfnamefont {K.}~\bibnamefont {Kaski}},\ }\bibfield
	{title} {\enquote {\bibinfo {title} {Dynamics of polymer ejection from
				capsid},}\ }\href {https://doi.org/10.1103/physreve.89.052702} {\bibfield
		{journal} {\bibinfo  {journal} {Phys. Rev. E}\ }\textbf {\bibinfo {volume}
			{89}} (\bibinfo {year} {2014})}\BibitemShut {NoStop}%
	\bibitem [{\citenamefont {Polson}(2015)}]{polson2015polymer}%
	\BibitemOpen
	\bibfield  {author} {\bibinfo {author} {\bibfnamefont {James~M}\ \bibnamefont
			{Polson}},\ }\bibfield  {title} {\enquote {\bibinfo {title} {Polymer
				translocation into and out of an ellipsoidal cavity},}\ }\href@noop {}
	{\bibfield  {journal} {\bibinfo  {journal} {J. Chem. Phys.}\ }\textbf
		{\bibinfo {volume} {142}},\ \bibinfo {pages} {174903} (\bibinfo {year}
		{2015})}\BibitemShut {NoStop}%
	\bibitem [{\citenamefont {Polson}\ and\ \citenamefont
		{Heckbert}(2019)}]{polson2019polymer}%
	\BibitemOpen
	\bibfield  {author} {\bibinfo {author} {\bibfnamefont {James~M.}\
			\bibnamefont {Polson}}\ and\ \bibinfo {author} {\bibfnamefont {David~R.}\
			\bibnamefont {Heckbert}},\ }\bibfield  {title} {\enquote {\bibinfo {title}
			{Polymer translocation into cavities: Effects of confinement geometry,
				crowding, and bending rigidity on the free energy},}\ }\href
	{https://doi.org/10.1103/physreve.100.012504} {\bibfield  {journal} {\bibinfo
			{journal} {Phys. Rev. E}\ }\textbf {\bibinfo {volume} {100}} (\bibinfo
		{year} {2019})}\BibitemShut {NoStop}%
	\bibitem [{\citenamefont {Ali}\ \emph {et~al.}(2008)\citenamefont {Ali},
		\citenamefont {Marenduzzo},\ and\ \citenamefont {Yeomans}}]{ali2008ejection}%
	\BibitemOpen
	\bibfield  {author} {\bibinfo {author} {\bibfnamefont {I.}~\bibnamefont
			{Ali}}, \bibinfo {author} {\bibfnamefont {D.}~\bibnamefont {Marenduzzo}}, \
		and\ \bibinfo {author} {\bibfnamefont {J.M.}\ \bibnamefont {Yeomans}},\
	}\bibfield  {title} {\enquote {\bibinfo {title} {Ejection dynamics of
				polymeric chains from viral capsids: Effect of solvent quality},}\ }\href
	{\doibase 10.1529/biophysj.107.111963} {\bibfield  {journal} {\bibinfo
			{journal} {Biophys. J.}\ }\textbf {\bibinfo {volume} {94}},\ \bibinfo {pages}
		{4159--4164} (\bibinfo {year} {2008})}\BibitemShut {NoStop}%
	\bibitem [{\citenamefont {Piili}\ \emph {et~al.}(2017)\citenamefont {Piili},
		\citenamefont {Suhonen},\ and\ \citenamefont {Linna}}]{piili2017uniform}%
	\BibitemOpen
	\bibfield  {author} {\bibinfo {author} {\bibfnamefont {J.}~\bibnamefont
			{Piili}}, \bibinfo {author} {\bibfnamefont {P.~M.}\ \bibnamefont {Suhonen}},
		\ and\ \bibinfo {author} {\bibfnamefont {R.~P.}\ \bibnamefont {Linna}},\
	}\bibfield  {title} {\enquote {\bibinfo {title} {Uniform description of
				polymer ejection dynamics from capsid with and without hydrodynamics},}\
	}\href {https://doi.org/10.1103/physreve.95.052418} {\bibfield  {journal}
		{\bibinfo  {journal} {Phys. Rev. E}\ }\textbf {\bibinfo {volume} {95}}
		(\bibinfo {year} {2017})}\BibitemShut {NoStop}%
	\bibitem [{\citenamefont {Linna}\ \emph {et~al.}(2017)\citenamefont {Linna},
		\citenamefont {Suhonen},\ and\ \citenamefont {Piili}}]{linna2017rigidity}%
	\BibitemOpen
	\bibfield  {author} {\bibinfo {author} {\bibfnamefont {R.~P.}\ \bibnamefont
			{Linna}}, \bibinfo {author} {\bibfnamefont {P.~M.}\ \bibnamefont {Suhonen}},
		\ and\ \bibinfo {author} {\bibfnamefont {J.}~\bibnamefont {Piili}},\
	}\bibfield  {title} {\enquote {\bibinfo {title} {Rigidity-induced scale
				invariance in polymer ejection from capsid},}\ }\href
	{https://doi.org/10.1103/physreve.96.052402} {\bibfield  {journal} {\bibinfo
			{journal} {Phys. Rev. E}\ }\textbf {\bibinfo {volume} {96}} (\bibinfo {year}
		{2017})}\BibitemShut {NoStop}%
	\bibitem [{\citenamefont {Plimpton}(1995)}]{plimpton1995fast}%
	\BibitemOpen
	\bibfield  {author} {\bibinfo {author} {\bibfnamefont {Steve}\ \bibnamefont
			{Plimpton}},\ }\bibfield  {title} {\enquote {\bibinfo {title} {Fast parallel
				algorithms for short-range molecular dynamics},}\ }\href@noop {} {\bibfield
		{journal} {\bibinfo  {journal} {J. Comput. Phys.}\ }\textbf {\bibinfo
			{volume} {117}},\ \bibinfo {pages} {1} (\bibinfo {year} {1995})},\ \bibinfo
	{note} {refer also to LAMMPS website
		(\url{http://lammps.sandia.gov/})}\BibitemShut {NoStop}%
	\bibitem [{\citenamefont {Weeks}\ \emph {et~al.}(1971)\citenamefont {Weeks},
		\citenamefont {Chandler},\ and\ \citenamefont {Andersen}}]{weeks1971role}%
	\BibitemOpen
	\bibfield  {author} {\bibinfo {author} {\bibfnamefont {John~D}\ \bibnamefont
			{Weeks}}, \bibinfo {author} {\bibfnamefont {David}\ \bibnamefont {Chandler}},
		\ and\ \bibinfo {author} {\bibfnamefont {Hans~C}\ \bibnamefont {Andersen}},\
	}\bibfield  {title} {\enquote {\bibinfo {title} {Role of repulsive forces in
				determining the equilibrium structure of simple liquids},}\ }\href@noop {}
	{\bibfield  {journal} {\bibinfo  {journal} {J. Chem. Phys.}\ }\textbf
		{\bibinfo {volume} {54}},\ \bibinfo {pages} {5237--5247} (\bibinfo {year}
		{1971})}\BibitemShut {NoStop}%
	\bibitem [{\citenamefont {Eisenriegler}\ \emph {et~al.}(1982)\citenamefont
		{Eisenriegler}, \citenamefont {Kremer},\ and\ \citenamefont
		{Binder}}]{eisenriegler1982adsorption}%
	\BibitemOpen
	\bibfield  {author} {\bibinfo {author} {\bibfnamefont {E.}~\bibnamefont
			{Eisenriegler}}, \bibinfo {author} {\bibfnamefont {K.}~\bibnamefont
			{Kremer}}, \ and\ \bibinfo {author} {\bibfnamefont {K.}~\bibnamefont
			{Binder}},\ }\bibfield  {title} {\enquote {\bibinfo {title} {Adsorption of
				polymer chains at surfaces: Scaling and monte carlo analyses},}\ }\href
	{\doibase 10.1063/1.443835} {\bibfield  {journal} {\bibinfo  {journal} {J.
				Chem. Phys.}\ }\textbf {\bibinfo {volume} {77}},\ \bibinfo {pages}
		{6296--6320} (\bibinfo {year} {1982})}\BibitemShut {NoStop}%
	\bibitem [{\citenamefont {Muthukumar}(2003)}]{muthukumar2003polymer}%
	\BibitemOpen
	\bibfield  {author} {\bibinfo {author} {\bibfnamefont {M.}~\bibnamefont
			{Muthukumar}},\ }\bibfield  {title} {\enquote {\bibinfo {title} {Polymer
				escape through a nanopore},}\ }\href {\doibase 10.1063/1.1553753} {\bibfield
		{journal} {\bibinfo  {journal} {J. Chem. Phys.}\ }\textbf {\bibinfo {volume}
			{118}},\ \bibinfo {pages} {5174--5184} (\bibinfo {year} {2003})}\BibitemShut
	{NoStop}%
	\bibitem [{\citenamefont {Bates}\ and\ \citenamefont
		{Watts}(1988)}]{bates1988nonlinear}%
	\BibitemOpen
	\bibfield  {author} {\bibinfo {author} {\bibfnamefont {D.~M.}\ \bibnamefont
			{Bates}}\ and\ \bibinfo {author} {\bibfnamefont {D.~G.}\ \bibnamefont
			{Watts}},\ }\href@noop {} {\emph {\bibinfo {title} {Nonlinear Regression
				Analysis and Its Applications}}}\ (\bibinfo  {publisher} {Wiley},\ \bibinfo
	{year} {1988})\BibitemShut {NoStop}%
	\bibitem [{\citenamefont {Kramers}(1940)}]{kramers1940brownian}%
	\BibitemOpen
	\bibfield  {author} {\bibinfo {author} {\bibfnamefont {H.A.}\ \bibnamefont
			{Kramers}},\ }\bibfield  {title} {\enquote {\bibinfo {title} {Brownian motion
				in a field of force and the diffusion model of chemical reactions},}\ }\href
	{\doibase 10.1016/s0031-8914(40)90098-2} {\bibfield  {journal} {\bibinfo
			{journal} {Physica}\ }\textbf {\bibinfo {volume} {7}},\ \bibinfo {pages}
		{284--304} (\bibinfo {year} {1940})}\BibitemShut {NoStop}%
	\bibitem [{\citenamefont {Gardiner}(2004)}]{gardiner2004handbook}%
	\BibitemOpen
	\bibfield  {author} {\bibinfo {author} {\bibfnamefont {Crispin~W}\
			\bibnamefont {Gardiner}},\ }\href@noop {} {\emph {\bibinfo {title} {Handbook
				of stochastic methods, 3rd ed.}}}\ (\bibinfo  {publisher} {Springer Berlin},\
	\bibinfo {year} {2004})\BibitemShut {NoStop}%
	\bibitem [{\citenamefont {Paturej}\ \emph {et~al.}(2019)\citenamefont
		{Paturej}, \citenamefont {Sommer},\ and\ \citenamefont
		{Kreer}}]{paturej2019universal}%
	\BibitemOpen
	\bibfield  {author} {\bibinfo {author} {\bibfnamefont {Jaros{\l}aw}\
			\bibnamefont {Paturej}}, \bibinfo {author} {\bibfnamefont {Jens-Uwe}\
			\bibnamefont {Sommer}}, \ and\ \bibinfo {author} {\bibfnamefont {Torsten}\
			\bibnamefont {Kreer}},\ }\bibfield  {title} {\enquote {\bibinfo {title}
			{Universal equation of state for flexible polymers beyond the semidilute
				regime},}\ }\href {https://doi.org/10.1103/physrevlett.122.087801} {\bibfield
		{journal} {\bibinfo  {journal} {Phys. Rev. Lett.}\ }\textbf {\bibinfo
			{volume} {122}} (\bibinfo {year} {2019})}\BibitemShut {NoStop}%
\end{thebibliography}

%merlin.mbs apsrev4-1.bst 2010-07-25 4.21a (PWD, AO, DPC) hacked
%Control: key (0)
%Control: author (0) dotless jnrlst
%Control: editor formatted (1) identically to author
%Control: production of article title (0) allowed
%Control: page (1) range
%Control: year (0) verbatim
%Control: production of eprint (0) enabled
%

\end{document}